\documentstyle[prd,aps,floats,amsfonts,amssymb,amsmath]{revtex}

\begin{document}
 \renewcommand{\theequation}{\thesection.\arabic{equation}}
 \draft
  \title{Cerenkov's Effect and Neutrino Oscillations in Loop Quantum Gravity}
\author{G. Lambiase$^{a,b}$\thanks{E-mail: lambiase@sa.infn.it}}
\address{$^a$Dipartimento di Fisica "E.R. Caianiello"
 Universit\'a di Salerno, 84081 Baronissi (Sa), Italy.}
  \address{$^b$INFN - Gruppo Collegato di Salerno, Italy.}
\date{\today}
\maketitle
\begin{abstract}
Bounds on the scale parameter ${\cal L}$ arising in loop quantum
gravity theory are derived in the framework of Cerenkov's effect
and neutrino oscillations. Assuming that ${\cal L}$ is an
universal constant, we infer ${\cal L}\gtrsim 10^{-18}$eV$^{-1}$,
a bound compatible with ones inferred in different physical
context.
\end{abstract}

\section{Introduction}
\setcounter{equation}{0}

The difficulties to build up a complete theory of quantum gravity
has motivated the development of semiclassical approaches in which
a Lorentz invariance breakdown occurs at the effective theory
level. The deformations of the Lorentz invariance manifest by
means of a slight deviation from the standard dispersion relations
of particles propagating in the vacuum. Such modifications has
been proposed in the paper \cite{amelino}, and derived in two
different approaches: String Theory \cite{ellis} and Loop Quantum
Gravity \cite{gambini,alfaroPRL,alfaroPRD}.

In this work we shall refer to the Alfaro, Morales-T\'ecotl,
Urrutia (AMU) paper \cite{alfaroPRL,alfaroPRD} which is based on
the loop quantum gravity \cite{rovelli}. In this model a new
length scale ${\cal L}$ occurs, with ${\cal L}\gg l_{Pl}\sim
10^{-19}$GeV$^{-1}$. The scale ${\cal L}$ separates the distances
$d$ that manifest the quantum loop structure of space ($d\ll {\cal
L}$) from the continuous flat space ($d\gtrsim {\cal L}$). As
observed in \cite{AlfaroPalma}, this length scale might give rise
to some observable effect of quantum gravity.

In the AMU formalism, the dispersion relation of freely
propagating fermions and photons have the form
\begin{equation}\label{disp}
  E_i^2=\tilde{A}_i^2 p_i^2+m_i^2\,,
\end{equation}
where $\tilde{A}_i$ encodes the Lorentz invariance deformation
${\cal L}$-dependent and is different for different species of
particles. More specifically, for Majorana fermions
\cite{alfaroPRL}, the dispersion relation is given by
\begin{equation}\label{disp-ferm}
  E_\pm^2=A_p^2p^2+\eta p^4 \pm 2\lambda p +m^2\,,
\end{equation}
where
\begin{equation}\label{coeff-ferm}
  A_p=1+\kappa_1\left(\frac{l_{Pl}}{{\cal L}}\right)+\kappa_2\left(\frac{l_{Pl}}{{\cal
  L}}\right)^2\,, \qquad \eta=\kappa_3l_{Pl}^2\,, \qquad
  \lambda=\kappa_5 \frac{l_{Pl}}{2{\cal L}^2}\,.
\end{equation}
The $\pm$ signs stand for the helicity of the propagating
fermions, and $\kappa_i$ are unknown coefficients of the order
$\kappa_i\sim {\cal O}(1)$.

For photons, the dispersion relation derived in AMU approach is
\cite{alfaroPRD}
\begin{equation}\label{disp-phot}
  \omega_\pm^2=k^2[A_\gamma^2\pm 2\theta_\gamma l_{Pl}k]\,,
\end{equation}
where
\begin{equation}\label{coeff-phot}
  A_\gamma=1+\kappa_\gamma\left(\frac{l_{Pl}}{{\cal
  L}}\right)^{2+2Y}\,.
\end{equation}
Again $\pm$ signs stand for the helicity dependence of photons in
the dispersion relation, $\kappa_\gamma \sim {\cal O}(1)$,
$\theta_\gamma \sim {\cal O}(1)$, and $Y = -1/2, 0, 1/2, 1,
\ldots$. It is worth note that a similar result has been obtained
by Gambini and Pullin \cite{gambini} with $A_\gamma=1$ and Ellis
et al. \cite{ellis} with the difference that the helicity
dependence is absent.

The aim of this paper is to determine bounds on the scale length
${\cal L}$. Alfaro and Palma \cite{AlfaroPalma} applies the AMU
theory to the observed Greizen-Zatsepin-Kuz'min (GZK) limit
anomaly and to the so called TeV-$\gamma$ paradox, i.e. the
detecton of high-energy photons with  a spectrum ranging up to 24
TeV from Mk 501, a BL Lac object at a red-shift of 0.034 ($\sim
134$Mpc). Assuming that no anomalies there exist, as recent works
seem to indicate \cite{stecker}, they find that, with a specific
choice of the $\kappa_i$-parameters, the favorite range for ${\cal
L}$ is $(Y=0)$
\begin{equation}\label{boundAP}
 4.6 \times 10^{-17}\mbox{eV}^{-1} \gtrsim {\cal L} \gtrsim 8.3\times
 10^{-18}\mbox{eV}^{-1}\,.
\end{equation}
Bounds on ${\cal L}$ are inferred in two different contexts: $i)$
the emission of radiation by charged particles via Cerenkov
effect; $ii)$ neutrino oscillations. The analysis is carried out
for ${\cal L}$ universal constant, and $k_3=0$ and $k_5=0$ in
(\ref{disp}). Besides, to explore the leading order helicity
independent effects of photons, we also put $\theta_\gamma=0$. A
detailed study including the helicity term in (\ref{disp-phot})
has been recently performed by Jacobson et al. in \cite{jacobson}.

\section{Cerenkov Effect in the AMU Theory}
\setcounter{equation}{0} This Section is devoted to study the
emission of the Cerenkov radiation by a charged particle whose
dispersion relation is modified by the microscopic loop structure
of space. As well known, the Cerenkov effect causes the charged
particles to decelerate radiatively, and the spectrum of the
radiated energy might be used as a sensitive means for probing
their properties. This effect has been also studied in
\cite{jacobson,major}, but the method to derive bounds on ${\cal
L}$ is different.

The dispersion relation of photons (\ref{disp-phot}) allows to
infer the phase velocity $v_\gamma$
\begin{equation}\label{v-phase}
 v_\gamma\equiv \frac{\omega}{k}\sim
 1+ (A_\gamma-1)\,.
\end{equation}
For massive particles, the groups velocity $v_p$, corresponding to
the physical velocity of the particles, is obtained by deriving
the energy (\ref{disp-ferm}) with respect to the momentum $p$
\begin{equation}\label{v-group}
 v_p\equiv \frac{dE}{dp}\sim
 1+(A_p-1)-\frac{m^2}{2p^2}\,.
\end{equation}
The ratio between the phase velocity (\ref{v-phase}) and the group
velocity (\ref{v-group}) is ($p \simeq E$)
\begin{equation}\label{ratio}
  \frac{v_\gamma}{v_p}=
  1+\frac{m^2}{2p^2}-(A_p-A_{\gamma})-(A_p-1)(A_\gamma-1) \,.
\end{equation}
The Cerenkov process occurs if the condition $v_\gamma/v_p<1$
holds, i.e.
\begin{equation}\label{cerenkov}
  \frac{m^2}{2p^2}-(1+A_pA_{\gamma}-2A_{\gamma})<0\,.
\end{equation}
In the limit ${\cal L}\to \infty$ the above relation reduces to
$m^2<0$, so that the Cerenkov effect is not kinematically allowed.

{} From Eq. (\ref{cerenkov}) one defines the threshold energy
$E_0$
\begin{equation}\label{E0}
  E_0=\frac{m}{\sqrt{2[1+A_{\gamma}(A_p-2)]}}\,,
\end{equation}
above which a charged particle is faster than the corresponding
phase velocity of light and must then emit Cerenkov radiation.

The rate of the emitted energy by the charged particle is
\cite{jackson}
\begin{eqnarray}\label{radiationemit}
  W(E)&=&\frac{dE}{dt}=e^2\int_0^{{\bar \omega}}\left(1-\frac{v_\gamma^2}{v_p^2}\right)\omega d\omega \\
  &=& e^2m^2\, \frac{(E^2-E_0^2)(E-E_0)^2}{2E_0^2E^2}\,, \nonumber
\end{eqnarray}
where the integration is overall the possible emitted frequencies
${\bar \omega}=E-E_0$ and Eqs. (\ref{ratio}) and (\ref{E0}) have
been used. The distance that charged particles can travel before
to relax their energy as Cerenkov radiation is
\begin{eqnarray}\label{travel}
  L(E_f)&=&\int_{x_f}^{x_i}dx=-\int_{E_i}^{E_f}\frac{dE}{W(E)} \\
         & \simeq & \frac{E_0}{4e^2m^2}\left\{ \ln\frac{(E_i-E_0)(E_f+E_0)}{(E_i+E_0)(E_f-E_0)}
         +6E_0\left(\frac{1}{E_f-E_0}-\frac{1}{E_i-E_0}\right)+2E_0^2\left[
         \frac{1}{(E_f-E_0)^2}-\frac{1}{(E_i-E_0)^2}\right]\right\} \nonumber
\end{eqnarray}
where $E_i$ denotes the initial energy of the particle, and $E_f$
is the final energy which differs from the threshold energy $E_0$
by an infinitesimal quantity, $E_f=(1+\zeta)E_0$, with $\zeta\ll
1$ ($E_i>E_f>E_0$). In the limit $E_i\gg E_0$, Eq. (\ref{travel})
reduces to $L(E_f)\sim E_0/2e^2m^2\zeta^2$. Thus the loop quantum
gravity effects become important on distance of cosmological
relevance, provided $\zeta$ is extremely small. In fact, for
$L(E_f)\sim $ Mpc and taking $E_0\sim 10^{9}$GeV as threshold
energy, it follows $\zeta< 10^{-8}$. Notice that $E_f=E_0$
corresponds to the asymptotic value $L(E_f)\to \infty$.

To determine a bound on the parameter ${\cal L}$, we shall work in
the context of cosmic rays physics. As widely believed, cosmic
rays have extragalactic origin \cite{11,12,13} with energies
exceeding $10^9$GeV, and sources located at distances greater than
$\sim$ few Mpc. Besides, at the energy greater than $10^{10}$GeV,
the spectrum should contain mostly protons \cite{12,15}. The large
number of observed events with energies exceeding $10^{10}$GeV
suggests that Cerenkov effect occurs with a threshold energy
expected to be of the order $E_0\gtrsim 10^{9}$GeV. For protons
with $m\sim$ GeV and for $Y=-1/2, 0$, we get

\begin{description}
  \item[-] $Y=-1/2$. In such a case $A_{\gamma}\sim 1+\kappa_\gamma
  (l_{Pl}/{\cal L})$, and we put $\kappa_2=0$ in the expression for
  $A_p\sim 1+\kappa_1(l_{Pl}/{\cal L})$. Being
  $\delta \kappa_1\equiv \kappa_1-\kappa_\gamma \sim {\cal O}(1)$ ($\kappa_1>\kappa_\gamma$),
  it follows
\begin{equation}\label{bound1}
  {\cal L}=\frac{2\delta \kappa_1 l_{Pl}E_0^2}{m^2}\gtrsim 10^{-8}\mbox{eV}^{-1}\,.
\end{equation}
This bound is nearly in the range of nuclear physics, so that it
can be discarded since no evidence of loop structure occurred at
this scale.

  \item[-] $Y=0$. The coefficient $A_\gamma$ reduces to  $A_\gamma\sim 1+\kappa_\gamma (l_{Pl}/{\cal
  L})^2$, and we put $\kappa_1=0$ thus
  $A_p\sim 1+\kappa_2(l_{Pl}/{\cal L})^2$.
  Being $\delta \kappa_2\equiv \kappa_2-\kappa_\gamma \sim {\cal
  O}(1)$ ($\kappa_2>\kappa_\gamma$), it follows
\begin{equation}\label{bound2}
  {\cal L}=\frac{\sqrt{2\delta \kappa_2}\,l_{Pl}E_0}{m}\gtrsim 10^{-18}\mbox{eV}^{-1}\,.
\end{equation}
This bound agrees with (\ref{boundAP}) derived in Ref.
\cite{AlfaroPalma}.

\end{description}

\section{Neutrino Oscillations in AMU Theory}
\setcounter{equation}{0}

The possibility that neutrinos particles may oscillate in
different flavor states is the most discussed problem of today
theoretical and experimental physics. Neutrino oscillations in the
vacuum occur owing to the non-degeneration of the mass-matrix
eigenvalues and to the difference of the mass eigenstates
$|\nu_1>$, $|\nu_2>$ from the weak interaction eigenstates
$|\nu_e>$, $|\nu_\mu>$. In the standard scenario, the energy
splitting between two mass eigenvalues $E_{iM}=m_i^2/2p$, $i=1,2$,
is $\Delta E=\Delta m^2/2p$, where $\Delta m^2=m_1^2-m_2^2$ is the
mass squared difference. The oscillation length is given
by\footnote{It is worth note that the dispersion relation can be
also modified by taking into account the effects of gravitational
fields, which might affect the quantum mechanical phase of massive
neutrinos (see for example \cite{DVA,konno,wudka}, and also
\cite{altri}).} $L_M=2\pi/\Delta E$.

In AMU's theory, the energy of particles is modified according to
Eq. (\ref{disp-ferm}):
\begin{equation}\label{disp-ferm-i}
  E_i=p+(A_p^i-1)p+\frac{m_i^2}{2p}\,,\quad i=1,2\,,
\end{equation}
where
\begin{equation}\label{coeff-A-i}
  A_p^i=1+\kappa_1^{(i)}\frac{l_{Pl}}{{\cal
  L}}+\kappa_2^{(i)}\left(\frac{l_{Pl}}{{\cal L}}\right)^2\,.
\end{equation}
The parameters $\kappa_j$, $j=1,2,\ldots$ are different for all
particles. It is hence natural to assume that $\kappa_1$ and
$\kappa_2$ are flavor depending. In this new setting, the energy
splitting is
\begin{equation}\label{split-en-AMU}
  \Delta E\approx \frac{\Delta m^2}{2p}+\delta A_p p\,,
\end{equation}
in which $\delta A_p=A_p^{(1)}-A_p^{(2)}$. The corresponding
oscillation length is therefore
\begin{equation}\label{osc-length-AMU}
  L=\frac{2\pi}{\displaystyle{\frac{\Delta m^2}{2p}}+\delta
  A_p p}=\frac{L_ML_{LQG}}{L_M+L_{LQG}}\,.
\end{equation}
Here, $L_{LQG}=2\pi/\delta A_p p$ is the oscillation length
induced by loop quantum gravity corrections. It is worth noting
that in Ref. \cite{alfaroPRL} the oscillation length of neutrinos
is $\sim p^{-2} l_{Pl}^{-1}$, a result obtained by imposing ${\cal
L}\sim p^{-1}$ (see also the recent papers \cite{hugo,foffa}). We
shall take a different point of view in which the scale length
${\cal L}$ is a free (and universal) parameter. The transition
probability is
\begin{equation}\label{prob}
P_{\nu_e\to \nu_\mu}=\sin^2 2\vartheta \sin^2\left(\frac{\pi
\Delta r}{L}\right)\,,
\end{equation}
being $\Delta r$ the distance travelled by neutrinos and
$\vartheta$ the vacuum mixing angle. Loop quantum gravity
corrections are relevant when $L_M\approx L_{LQG}$, i.e.
\begin{equation}\label{deltaA}
  \delta A_p \approx \frac{\Delta m^2}{2p^2}\,.
\end{equation}
Let us discuss separately the case $\kappa_2=0$ and $\kappa_1=0$.

\begin{description}
  \item[-] $\kappa_2=0$. Denoting with
  $\delta\kappa_1=\kappa_1^{(1)}-\kappa_1^{(2)}\sim {\cal O}(1)$,
  and using Eq. (\ref{coeff-A-i}), one gets
\begin{equation}\label{bound-neutrino}
  {\cal L}=\frac{2l_{Pl}\delta\kappa_1 p^2}{\Delta m^2}\,.
\end{equation}
For solar neutrinos, $\Delta m^2\sim 10^{-10}$eV$^2$ and $p\sim $
MeV, so that
\begin{equation}\label{bound1-solar}
  {\cal L}\sim 10^{-6}\mbox{eV}^{-1}\,.
\end{equation}
In the case of atmospheric neutrinos, $\Delta m^2\sim
10^{-3}$eV$^2$ and $p\sim $GeV, which implies
\begin{equation}\label{bound1-atmosph}
  {\cal L}\sim 10^{-7}\mbox{eV}^{-1}\,.
\end{equation}
According to previous discussion, these values have to be
discarded since no evidence of loop structure of space has been
observed at this scale.

  \item[-] $\kappa_1=0$. By defining
  $\delta\kappa_2=\kappa_2^{(1)}-\kappa_2^{(2)}\sim {\cal O}(1)$,
  and using Eq. (\ref{coeff-A-i}), one obtains
\begin{equation}\label{bound-neutrino2}
  {\cal L}=\frac{\sqrt{2\delta\kappa_2 }\,l_{Pl}p}{\sqrt{\Delta m^2}}\,.
\end{equation}
For solar neutrinos it follows
\begin{equation}\label{bound2-solar}
  {\cal L}\sim 10^{-18}\mbox{eV}^{-1}\,.
\end{equation}
whereas for atmospheric neutrinos one gets
\begin{equation}\label{bound2-atmosph}
  {\cal L}\sim  10^{-18}\mbox{eV}^{-1}\,.
\end{equation}
In both cases, the scale length ${\cal L}$ belongs to the bound
derived by Alfaro and Palma (\ref{boundAP}).
\end{description}
We conclude noting that loop quantum gravity provides us a scheme
in which flavor oscillations can occur even for massless particles
or for massive but degenerate neutrinos.

\section{Conclusion}
Despite the concrete difficulties to probe quantum gravity effects
occurring at the Planck scale, recently there has been an increase
in the conviction that quantum gravity should predict a slight
departure from the Lorentz invariance, which manifests itself in a
deformation of the dispersion relations of photons and fermions.
Such results have been indeed inferred in loop quantum gravity
\cite{gambini,alfaroPRL,alfaroPRD} and string theory \cite{ellis}.
In both approaches, a scale length, characterizing the scale on
which new effects are non trivial, appears. The natural arena for
probing the occurrence of the new scale is provided by gamma ray
bursts physics, and by the observed GZK limit anomaly and related
processes \cite{AlfaroPalma} (see also Refs.
\cite{Sudarsky,neweffects} for different contexts).

The aim of this paper has been to infer bounds on the length scale
${\cal L}$ occurring in the AMU theory. Such bounds have been
determined in the context of Cerenkov effect and neutrinos
oscillations. Our analysis yields
 \[
 {\cal L}\gtrsim 10^{-18}\mbox{eV}^{-1}\,.
 \]
Remarkably, it is compatible with bounds determined in Ref.
\cite{AlfaroPalma}. Thus, corrections induced by loop structure of
space becomes relevant for distances bigger than
$10^{-18}$eV$^{-1}$, an estimation that gives some hope to bring
quantum gravity effects to the realm of experimental
results\footnote{To estimate the order of magnitude of ${\cal L}$
is certainly of current interest, and, in fact, many attempts have
been proposed in different systems. A stringent bound on
$k_1l_{Pl}/{\cal L}$ has been recently inferred by
Sudarsky,Urrutia and Vucetich in Ref. \cite{Sudarsky}. Using their
notation, we have
 \[
 k_1l_{Pl}/{\cal L}\to \Theta_1 l_{Pl}m
 \]
with $\Theta_1<10^{-5}$. Such a bound has been derived for the
mobile scale, i.e. ${\cal L}$ is fixed by the mass of particles
${\cal L}\sim 1/m$ (the nucleon mass in the case of Ref.
\cite{Sudarsky}). Results of Ref. \cite{Sudarsky} seem to be very
promising to probe quantum gravity effects as following from AMU
theory.}.


\acknowledgments It is a great pleasure for the author to thank J.
Alfaro, H. Morales-T\'ecolt, and L.F. Urrutia for reading the
paper. Many thanks also to D.V. Ahluwalia for discussions on
Cerenkov's effect during his visit in Salerno, and to the referee
of MPLA for comments.

\end{document}